\documentclass[reprint,amsmath,amssymb,aps,pre,]{revtex4-2}
\usepackage[caption=false]{subfig}
\usepackage{listings}
\usepackage{xcolor}
\usepackage{multirow}
\usepackage{appendix}
\usepackage{graphicx}
\usepackage{adjustbox}
\usepackage{dcolumn}
\usepackage{bm}
\usepackage{hyperref}
\usepackage{braket}
\usepackage{url}
\makeatletter
\newcommand{\vast}{\bBigg@{3}}
\newcommand{\Vast}{\bBigg@{4}}
\begin{document}
\title{Vaccination Dilemma in the thermodynamic limit}
\author{Colin Benjamin}
\email{colin.nano@gmail.com}
\author{Arjun Krishnan U M}
\affiliation{School of Physical Sciences, National Institute of Science Education and Research Bhubaneswar, Jatni 752050, India}
\affiliation{Homi Bhabha National Institute, Training School Complex, Anushaktinagar, Mumbai 400094, India}
\date{\today}
\begin{abstract}
The vaccination game is a social dilemma that refers to the conundrum individuals face (to get immunized or not) when the population is exposed to an infectious disease. The model has recently gained much traction due to the COVID-19 pandemic since the public perception of vaccines plays a significant role in disease dynamics. This paper studies the vaccination game in the thermodynamic limit with an analytical method derived from the $1D$ Ising model called Nash equilibrium mapping. The individual dilemma regarding Vaccination comes from an internal conflict wherein one tries to balance the perceived advantages of immunizing with the apparent risks associated with Vaccination which they hear through different news media. We compare the results of Nash equilibrium(NE) mapping from other $1D$ Ising-based models, namely Darwinian evolution and agent-based simulation. This study aims to analyze the behavior of an infinite population regarding what fraction of people choose to vaccinate or not vaccinate. While Nash equilibrium mapping and agent-based simulation agree mostly, Darwinian evolution strays far from the two models. It fails to predict the equilibrium behavior of players in the population reasonably. We apply the results of our study to analyze the Astra-Zeneca(AZ) COVID-19 vaccine risk versus disease deaths debate, both via NE mapping and agent-based method. Both predict near $100\%$ AZ vaccine coverage for people aged above $40$, notwithstanding the risk. At the same time, younger people show a slight reluctance. We predict that while government intervention via vaccination mandates and/or advertisement campaigns is unnecessary for the older population, for the younger population (ages: $20-39$), some encouragement from the government via media campaigns and/or vaccine mandates may be necessary.
\end{abstract}
\maketitle
\textbf{In this work, we study the vaccination dilemma game confronted by individuals in a population during the spread of an infectious disease. We approach this problem by utilizing Nash equilibrium mapping- an analytical method based on the 1D Ising model used to study the equilibrium behavior of players in a game in the thermodynamic limit. The vaccination game was introduced in [Chris T. Bauch, David J. D. Earn, PNAS September 7,2004 101 (36) 13391-13394] to study how the perception of risks associated with vaccine shots and disease contraction affect the vaccine coverage among a population. Our work attempts to represent this game as a 2-group, 2-strategy pairwise interaction normal form game and uses the Nash equilibrium mapping model to find results in the thermodynamic limit. Our results are further compared against numerical agent-based simulation results to establish their validity. Additionally, we attempt to make projections of the Astra-Zeneca vaccine coverage among different age groups in the United Kingdom by using data from surveys to estimate the potential risk (reported cases of blood clots) and benefits (prevented ICU admissions) of the vaccine as well as the cost of disease contraction (deaths from COVID-19). Our model also brings in a metric for government mandates and media promotion affecting vaccine coverage.}
\section{Introduction}
Epidemiologists and mathematicians have continuously developed concise mathematical models to study the spread of infectious diseases in a population. Since the advent of the COVID-19 pandemic, the field has gained much traction. Game theory especially has gained much attention since the perception of a general populace towards the consequence of contracting the disease and their disposition towards preventive measures (or vaccines) can be described as a game. How this perception changes over time plays a significant role in formulating a definitive plan to get the pandemic under control. Another critical problem is how a population feels about vaccine administration once it has been developed. Due to widespread misinformation on social media about the side effects of vaccines, a not insignificant proportion of the population may decide to hold off on immunization. Ref.~\cite{Bauch} introduced the vaccination game, which attempts to study this problem by combining game theory dynamics with an epidemiological SIR compartmental model. This work looks at the thermodynamic or infinite-player limit of the Vaccination dilemma/game using Nash equilibrium mapping (NEM). NEM is an analytical method used to study games in the infinite player limit via mapping to the 1D Ising model.

Such a formulation allows us to study the behavior of a large and complex population. When it comes to the vaccination game, it measures what fraction of the population choose to (or not to) take vaccine shots. The vaccination game, as introduced in Ref.~\cite{bauch,Fu} is meant for a well-mixed population where every individual interacts with each other. However, the NE mapping model allows for pairwise interactions only between nearest neighbors. We proceed by constructing a 2-player 2-strategy payoff matrix from the payoffs associated with choosing to vaccinate or not. The game magnetization and average-payoff/player(APP) in the zero-noise limit obtained via NEM are compared with ABM simulations. Additionally, we also compare it with an alternate analytical model- again derived via mapping to the 1D Ising model- named Darwinian evolution. While NEM and ABM simulations seem to agree for most of the payoff domain reasonably, Darwinian evolution differs wildly from the two models. It gives results incompatible with the Nash equilibrium of the game. Further, we will use these methods to analyze survey data of potential benefits and harms of the Astra-Zeneca COVID-19 vaccine and how it affects the coverage among different age groups of a population.

In the next section, we will cover the theoretical background for the NE mapping and agent-based simulations. Later on, we go on to study how these methods are applied to the vaccination dilemma. The final section will discuss our results and how the analytical methods and numerical agent-based simulation results compare. The paper ends with a conclusion and an appendix wherein we approach the same problem except using the framework of the Darwinian evolution method, which is also based on the $1D$ Ising model.
\section{Theory}
In this section, we will define the Vaccination game for a 2-group, 2-strategy system. We also explain the NEM method in detail, which utilizes the mathematics of equilibrium statistical physics through an analogy with the $1D$ Ising model in the thermodynamic limit to study social dilemmas with an infinite number of players. Further, we elaborate on ABM simulations used to mimic a system of interacting players in a game and obtain the equilibrium state.
\subsection{\label{vaccinationdilemma theory}Vaccination game}
In the vaccination game as introduced in Ref.~\cite{Bauch}, an individual in a well-mixed population chooses to vaccinate with probability $P$. The vaccine in this model is assumed to give perfect immunity towards the disease spreading among the population. $r_{v}$ is the anticipated vaccination cost or risk of Vaccination. $r_{s}$ is the cost of contracting the disease, which is assumed to be constant among the whole population, and $\pi_p$ is the risk of a non-vaccinated individual contracting the illness $p$ is the fraction of the population that is vaccinated. The perceived payoffs from vaccinating or non-vaccinating is given as follows,
\begin{equation}
E(P, p)=P\left(-r_{v}\right)+(1-P)\left(-r_{s} \pi_{p}\right)
\end{equation}
We need to convert this formulation into one which is suitable for pairwise interaction models we mean to study. In order to do this, we assume the population is made of two groups $A$ and $B$. When both groups $A$ and $B$ vaccinate, they earn a payoff $-r_{v}$. When only one of the groups is vaccinated and the other chooses not to, then $p=0.5$ since half the population is vaccinated. In this case payoff for vaccinated individuals is $-r_{v}$ and for non-vaccinated, payoff is $-\pi_{0.5}r_s$. When both groups $A$ and $B$ choose not to vaccinate, $p=0$, and each group earns a payoff: $-\pi_{0}r_s$. Thus we get the 2-group, 2-strategy payoff matrix as,
\begin{equation}
\mathcal{U}=
\left(\begin{array}{c|c c}
& V & NV \\
\hline
V & -r_v,-r_v & -r_v,-\pi_{0.5}r_s \\
NV & -\pi_{0.5}r_s,-r_v & -\pi_{0}r_s,-\pi_{0}r_s
\end{array}\right).
\label{matrix1}
\end{equation}
Here, $V$ and $NV$ stand for vaccinate and non-vaccinate strategies respectively. The formula for risk of a non-vaccinated individual contracting the illness $\pi_p$ when $p$ fraction of population is vaccinated was obtained in Ref.~\cite{Bauch} as,
\begin{equation}
\pi_p=1-\frac{1}{R_0(1-p)}.
\end{equation}
It is obtained from the SIR epidemiological model by taking the proportion of the unvaccinated and susceptible individuals who become infected in a unit of time. Here, $R_0$ is the basic reproductive ratio of the infection. $R_0$ for COVID-19 was estimated by the World Health Organization to be between 1.5 and 2.5 \cite{WHO}. Thus payoff matrix from Eq.~(\ref{matrix1}) reduces to:
\begin{equation}
\mathcal{U}=
\left(\begin{array}{c|c c}
& V & NV \\
\hline
V & -r_v,-r_v & -r_v,-(1-\frac{2}{R_0})r_s \\
NV & -(1-\frac{2}{R_0})r_s,-r_v & -(1-\frac{1}{R_0})r_s,-(1-\frac{1}{R_0})r_s
\end{array}\right).
\label{vaccinationmatrix}
\end{equation}
In Eq.~(\ref{vaccinationmatrix}), we assume that $r_v<r_s$ because if people believe that the cost of Vaccination is more than the cost of illness, they will not find any incentive to get vaccinated. There are three distinct payoff regimes for the Vaccination dilemma whose Nash equilibrium can be determined as follows:
\begin{enumerate}
\item When $r_v<r_s(1-\frac{2}{R_0})$, pure strategy Nash equilibrium is $(V,V)$. Here, the Pareto optimum of the game is also $(V, V)$.
\item When $r_s(1-\frac{2}{R_0})<r_v<r_s(1-\frac{1}{R_0})$, there are two pure strategy Nash equilibrium- $(V,NV)$ and $(NV,V)$. Additionally, a mixed strategy Nash equilibrium $(\sigma,\sigma)$ exists, with $\sigma$ being the case where a player chooses the $V$ strategy with probability $p*=R_0(1-\frac{1}{R_0}-\frac{r_v}{r_s})$ and $NV$ strategy with probability $(1-p*)$. In this payoff regime, the Pareto optimal outcomes are $(V,NV)$ and $(NV,V)$.
\item When $r_v>r_s(1-\frac{1}{R_0})$, pure strategy Nash equilibrium is $(NV,NV)$. Pareto optimum of the game is also $(NV,NV)$.
\end{enumerate}
Cases $1$ and $3$ are similar to the public goods games whose Nash equilibrium strategies are cooperation and defection, respectively. Here, each group ends with a lesser payoff if they decide not to adopt the Nash equilibrium strategy. Case $3$ is similar to the Hawk-Dove game with a mixed strategy Nash equilibrium and two pure strategies Nash equilibrium. In this regime, one group is better off when they decide not to vaccinate, given that the other group gets vaccinated.
\subsection{\label{Isingmodeltheory}1D Ising model}
We study $1D$ Ising model for our work since the analytical game theory models we use in this paper are derived by considering the analogy of player interactions with interaction between spins in a $1D$ chain. For an $N$ spin site system with coupling constant $J$ and external magnetic field $h$, the Hamiltonian is given by,
\begin{equation}
{\mathcal H}=-J\sum_{i=1}^N\tau_i\tau_{i+1}-h\sum_{i=1}^N\tau_i,
\label{eqn:NEHam}
\end{equation}
here $\tau=+1$ denotes up spin $\tau=-1$ denotes down spin. One can obtain the partition function($\mathcal{Z}$) from the above equation. Further the free energy of the system is calculated using $F=-k_BT\ln(\mathcal{Z})$. Magnetization (fraction of up spins minus fraction of down spins in the system) and free energy can be derived easily (see Ref.~\cite{Ising}),
\begin{equation}
m=-\frac{1}{N}\frac{dF}{dh}=\frac{1}{N}\frac{1}{\beta}\frac{1}{\mathcal{Z}}\frac{d\mathcal{Z}}{dh}=\frac{\sinh(\beta h)}{\sqrt{\sinh^2(\beta h)+e^{-4\beta J}}}.
\label{Isingmagnetisation}
\end{equation}
Here, $\beta=1/k_BT$.
\subsection{\label{NEtheory}Nash equilibrium mapping (NEM)}
We utilize NEM method in our work because the model allows us to quantitatively study the equilibrium behavior of players in a population. In order to understand NEM method for games in thermodynamic limit, we look at the $1D$ Ising model for $N=2$. The Hamiltonian of such a system with two spin sites is,
\begin{equation}
\mathcal{H}=-J(\tau_1\tau_2+\tau_2\tau_1)-h(\tau_1+\tau_2).
\end{equation}
Individual energies of spin site $1$ and spin site $2$ are,
\begin{equation}
\mathcal{E}_1=-J\tau_1\tau_2-h\tau_1\hspace{5mm}\text{ and }\hspace{5mm}\mathcal{E}_2=-J\tau_2\tau_1-h\tau_2.
\end{equation}
In social dilemmas, players are attracted to maximum payoffs, while the minimum energy defines equilibrium in physical systems. To make a comparison between payoff matrix and energy matrix, we frame this  such that spin sites try to minimize their energies $-\mathcal{E}_i$ concerning spins $\tau = +1,-1$ (See Refs.~\cite{galam, Colin, Colin2, Colin3}).
\begin{equation}
-\mathcal{E}= \Vast(\begin{array}{c|c c}
& \tau_2=+1 & \tau_2=-1 \\
\hline
\tau_1=+1 & J+h,J+h & -J+h,-J-h \\
\tau_1=-1 & -J-h,-J+h & J-h,J-h
\label{eqn:Isingmatrix}
\end{array}\Vast).
\end{equation}
To make the connection between payoff's and energies we write the payoff matrix for a symmetric two-player (or, group) social dilemma as,
\begin{equation}
\mathcal{U}=
\left(\begin{array}{c|c c}
& st_1 & st_2 \\
\hline
st_1 & a_{1},a_{1} & a_{2},a_{3} \\
st_2 & a_{3},a_{2} & a_{4},a_{4}
\end{array}\right).
\label{eqn:abcdmatrix}
\end{equation}
wherein $st_{1}, st_{2}$ are the strategies of the two players (or, groups) and $a_{1}, a_{2}, a_{3}, a_{4}$ are the payoff's corresponding to those strategies. To make the connection between Ising model spin sites and players in a game, we now introduce a transformation that doesn't affect the Nash equilibrium of the game\cite{galam,devos},
\begin{equation*}
\mathcal{U'}=
\left(\begin{array}{c|c c}
& st_1 & st_2 \\
\hline
st_1 & a_{1}+\gamma,a_{1}+\gamma & a_{2}+\delta,a_{3}+\gamma \\
st_2 & a_{3}+\gamma,a_{2}+\delta & a_{4}+\delta,a_{4}+\delta
\end{array}\right).
\end{equation*}
If $\gamma$ and $\delta$ take values as given below,
\begin{align}
\gamma=\frac{-(a_{1}+a_{3})}{2},\text{ } \delta=\frac{-(a_{2}+a_{4})}{2}.
\end{align}
then this gives us the transformed payoff matrix as,
\begin{equation}
\mathcal{U'}=
\Vast(\begin{array}{c|c c}
& st_1 & st_2 \\
\hline
st_1 & \frac{a_{1}-a_{3}}{2},\frac{a_{1}-a_{3}}{2} & \frac{a_{2}-a_{4}}{2},\frac{a_{3}-a_{1}}{2} \\
st_2 & \frac{a_{3}-a_{1}}{2},\frac{a_{2}-a_{4}}{2} & \frac{a_{4}-a_{2}}{2},\frac{a_{4}-a_{2}}{2}
\end{array}\Vast).
\label{eqn:transformedpayoffmatrix}
\end{equation}
The transformed payoff matrix is now structurally similar to that of~(\ref{eqn:Isingmatrix}). By equating elements of the two matrices, we get the following relation,
\begin{equation}
J=\frac{a_{1}-a_{3}+a_{4}-a_{2}}{4}\hspace{5mm} \text{ and }\hspace{5mm}h=\frac{a_{1}-a_{3}+a_{2}-a_{4}}{4}.
\label{eqn:Jandh}
\end{equation}
Substituting this in~(\ref{Isingmagnetisation}), we get game magnetization $m_g$ (i.e., difference between fraction of players acting with strategy $st_1$ vis-a-vis players acting with strategy $st_2$ in the social dilemma described by payoff matrix~(\ref{eqn:abcdmatrix})) in thermodynamic limit as,
\begingroup
\begin{equation}
m_g=\frac{\sinh \beta (\frac{a_{1}-a_{3}+a_{2}-a_{4}}{4})}{\sqrt{\ \sinh^2\beta( \frac{a_{1}-a_{3}+a_{2}-a_{4}}{4})+e^{-4\beta(\frac{a_{1}-a_{3}+a_{4}-a_{2}}{4})}}},
\label{eqn:magNEmappingabcdgeneral}
\end{equation}
\endgroup
herein, $\beta=1/(k_{B}T)$. The average-payoff obtained by each player in the social dilemma gives us additional information about distribution of strategies among individuals in the population. In order to derive this, we begin with the partition function of the $1D$ Ising chain with only two sites,
\begin{equation}
\mathcal{Z}=\sum_{\tau_1,\tau_2}e^{-\beta \mathcal{H}}=e^{\beta(2J+2h)}+2e^{-\beta(2J)}+e^{\beta(2J-2h)},
\label{H-2site,Z-2site}
\end{equation}
The average thermodynamic energy $\langle \mathcal{E} \rangle$ is then~\cite{Reif} ,
\begin{align*}
\langle \mathcal{E} \rangle=&-\frac{\partial \ln \mathcal{Z}}{\partial \beta}.
\label{eqn:averagethermodynamicenergy}
\end{align*}
Since we took the negative of energy values to get the payoff matrix in~(\ref{eqn:Isingmatrix}), the average-payoff/player(APP) in the social dilemma is given by $-\langle \mathcal{E} \rangle$. Further, since we started with the Hamiltonian corresponding to the sum of the energies of two sites, we divide by $2$ to get the APP: $\langle  \mathcal{U'} \rangle$, as
\begin{align}
&\langle \mathcal{U'} \rangle= -\frac{\langle \mathcal{E} \rangle}{2}=\frac{1}{2} \frac{\partial \ln \mathcal{Z}}{\partial \beta}=\\
&\frac{\resizebox{.4 \textwidth}{!} {$(J+h)e^{\beta (2J+2h)}-(2J)e^{-\beta (2J)}+(J-h)e^{\beta (2J-2h)}$}}{e^{\beta (2J+2h)}+2e^{-\beta (2J)}+e^{\beta (2J-2h)}}.
\label{averagepayoffgeneral}
\end{align}
Note that the APP is in terms of transformed payoff matrix in~(\ref{eqn:transformedpayoffmatrix}). Temperature in the context of game theory is a measure of noise or selection pressure. In the infinite temperature limit, strategy choices are completely randomized, and the players do not have any affinity towards a particular strategy. At zero temperature, i.e., zero-noise limit, there is the absence of any randomization in strategic choices. Hence, the population favors the Nash equilibrium strategy of the corresponding two-group social dilemma.
\subsubsection{\label{NEMapping:correct}Correctness of the NEM method}
Evolutionary game theory studies are more often than not are based on approaches using non-equilibrium dynamics. An approach using equilibrium statistical mechanics was pioneered by Ref.~\cite{Adami} wherein the authors introduced Hamiltonian dynamics(HD) and Darwinian evolution(DE) methods. One among us, intrigued by the novelty of this, introduced the NEM method which also uses equilibrium statistical mechanics and a mapping to  the 1D Ising model and showed that it gives correct results as compared to HD method, see Refs.~\cite{Colin,Colin2}, while the incorrectness of DE method vis-à-vis NEM method was proved in Ref.~\cite{NEvsDE}. In the appendix of this paper too, we find that the DE method fails to account properly for the Vaccination dilemma.  The  motivation for this paper is not to look at time dependent dynamics of evolution of strategies, but rather to invoke the mathematics of equilibrium statistical mechanics by drawing comparisons with solution of $1D$ Ising model. The NEM method applied to the Vaccination dilemma in question does not present a system of players that evolve their strategy over time. It rather presents a picture of a one-shot game, where infinite number of players decide their strategies in one go, while subject to  noise (or, temperature), defined via $\beta$.

\subsection{\label{ABMtheory}Agent based method (ABM)}
An agent-based simulation is a numerical method used in analyzing social dilemmas in the thermodynamic limit via simulating the interaction between players, similar to that done via a 1D Ising chain. We consider about $10,000$ sites (or players) configured as a $1D$ chain, with each site interacting with its immediate neighbor. Each site has two available strategies, i.e., either vaccinate or non-vaccinate. Each site's energy is dependent on its strategy and that of its nearest neighbor and is defined via the relation $\mathcal{E}=-\mathcal{U}$, i.e., the negative of the payoff matrix. Strategies are randomly updated around 10 million times, which is $1000$ updates per site in the $1D$ chain. The steps for the algorithm is given below:
\begin{enumerate}
\item All sites are assigned a random strategy, i.e., either $(vaccinated/non-vaccinated)$.
\item A random spin site $i$ is chosen, and its energy is $\mathcal{E}_i$ deduced based on its strategy and that of its nearest neighbor using the energy matrix $\mathcal{E}$.
\item $\Delta \mathcal{E}_i$ is the change in energy if the spin site $i$ had its strategy been flipped (i.e., $vaccinate$ to $non-vaccinate$ or vice versa) while the nearest neighbor strategy remains unchanged.
\item Strategy of site $i$ is flipped according to the Fermi transition probability $1/(1+e^{\beta \Delta \mathcal{E}_i})$ (See Refs.~\cite{altrock,bladon}).
\item Go over to step $2$ to pick another random spin site, and this is repeated about 10 million times.
\item Difference between the fraction of vaccinated and non-vaccinated in the system's final state is calculated to determine game magnetization $m_g$.
\end{enumerate}
The probability for a random spin site $i$ to flip its strategy increases as $\Delta \mathcal{E}_i$ decreases. Hence with each alteration in strategy, the system is heading towards the equilibrium state with minimum energy. The next section will go over the results of analytical NEM and  ABM simulations when applied to the vaccination dilemma.

\section{Results}
The game magnetization and APP as derived by NEM and ABM simulations for vaccination dilemma are described here. These give us an understanding of the strategies adopted by individuals in a population. We are also interested in the APP at zero temperature (or zero noise) limit, since the randomization in strategies is non-existent and allows for direct comparison with the payoff of Nash equilibrium of the game, hence it gives us a measure of the accuracy of the model in predicting the Nash equilibrium behavior of all players. Additionally, it allows us to verify the completely random behavior of the players at  infinite-noise (or the infinite temperature) limit.
\subsection{\label{Chicken-NE}Results from NE mapping}

\subsubsection{\label{Chicken-NE-mag}Game magnetization}
The game magnetization, i.e., difference between fraction of vaccinated and un-vaccinated population in the thermodynamic limit according to NE mapping method for the vaccination dilemma~(\ref{vaccinationmatrix}),(\ref{eqn:magNEmappingabcdgeneral}) is given by,
\begin{equation}
m_g=\frac{\sinh\beta\left(\frac{-2r_v+\left(2-\frac{3}{R_{0}}\right)r_{s}}{4}\right)}{\sqrt{\sinh^2\beta\left(\frac{-2r_v+\left(2-\frac{3}{R_{0}}\right)r_{s}}{4}\right)+e^{\frac{\beta r_{s}}{R_{0}}}}}.
\label{eqn:NEmagchicken}
\end{equation}
Game magnetization veers to zero in the infinite-noise ($T \xrightarrow{} \infty$ or $\beta \xrightarrow{} 0$) limit. This happens since noise randomizes the strategic choices of the players and for infinite-noise, players' strategic choices are completely random leading to equal population of vaccinated and non-vaccinated individuals and hence game magnetization vanishes. In case of zero noise limit ($T \xrightarrow{} 0$ or $\beta \xrightarrow{} \infty$), game magnetization is $+1$ for $r_v<(1-\frac{5}{2R_0})r_s$, $-1$ for $r_v>(1-\frac{1}{2R_0})r_s$ and zero for $(1-\frac{1}{2R_0})r_s<r_v<(1-\frac{5}{R_0})r_s$. We plot the game magnetization $m_g$ against vaccination cost $r_v$ at different $\beta$ values in Fig.~(\ref{fig:Magvsr1}). Here we can see how the game magnetization changes as the game goes from the payoff domain where Nash equilibrium strategy is vaccination, to the domain where there is mixed strategy Nash equilibrium (along with pure strategy Nash equilibrium $(V,NV)$ and $(NV,V)$) and finally to the domain where Nash equilibrium strategy is non-vaccination. We can see that regardless of the value of $\beta$, $m_g$ vanishes at $r_v=(1-\frac{3}{2R_0})r_s$ where mixed Nash equilibrium is given by both strategies being equi-probable. In order to get a better understanding, in the next subsection, we calculate the APP both in the nil or zero-noise as well as infinite-noise limits.
\subsubsection{\label{Chicken-NE-avgU}Average-payoff/player (APP)}
The transformed payoff matrix for the vaccination dilemma~(\ref{eqn:transformedpayoffmatrix}),
\begin{equation}
\mathcal{U'}=
\left(\begin{array}{c c}
\frac{(1-\frac{2}{R_0})r_s-r_v}{2} & \frac{(1-\frac{1}{R_0})r_s-r_v}{2} \\
-\frac{(1-\frac{2}{R_0})r_s-r_v}{2} & -\frac{(1-\frac{1}{R_0})r_s-r_v}{2}
\end{array}\right).
\label{transformedChickenmatrix}
\end{equation}
This is the payoff matrix of the one group and payoff of the second group can be deduced from the fact that the transformed 2-group vaccination dilemma matrix is symmetric. From Eq.~(\ref{averagepayoffgeneral}), $J$ and $h$ can be written in terms of payoffs~(\ref{eqn:Jandh}) as follows- $J=\frac{a_{1}+a_{4}-a_{2}-a_{3}}{4}=-\frac{r_s}{4R_0}$ and $h=\frac{a_{1}+a_{2}-a_{3}-a_{4}}{4}=\frac{-2r_v+(2-\frac{3}{R_0})r_s}{4}$. This gives us the APP as,
\begin{align}
\langle \mathcal{U'} \rangle= -\frac{\langle \mathcal{E} \rangle}{2}=\frac{xe^{2\beta x}+2ye^{\beta y}+ze^{2\beta z}}{e^{2\beta x}+2e^{\beta y}+e^{2\beta z}},
\label{vaccinationgameaveragepayoff}
\end{align}
where $x=(-r_v+r_s(1-\frac{1}{2R_0}))/2,y=r_s/2R_0$ and $z=(r_v-r_s(1-\frac{1}{R_0}))/2$. In the  infinite-noise (or, $\beta \xrightarrow{} 0$)  limit, we get the APP as,
\begin{equation}
\underset{\beta \xrightarrow{} 0}{lim} \langle \mathcal{U'} \rangle = 0.
\end{equation}
This is merely the average of the payoffs from \textit{(V,V), (V,NV), (NV,V)} and \textit{(NV,NV)} strategies as can be seen by taking the mean of all the payoffs from the transformed payoff matrix~(\ref{transformedChickenmatrix}). This is because all strategy pairs are equally probable in infinite-noise (or, $\beta \xrightarrow{} 0$) limit since strategic choices of players are completely randomized in this limit. When we represent the APP in infinite-noise limit in terms of real vaccination dilemma payoffs~(\ref{vaccinationmatrix}), we get,
\begin{equation}
\underset{\beta \xrightarrow{} 0}{lim} \langle \mathcal{U} \rangle = \frac{-2r_v-2r_s+3r_s/R_0}{4},
\label{zeroUlimit}
\end{equation}
which is obtained by taking mean of all the payoffs in the original vaccination dilemma payoff matrix. Next we are interested in finding the APP in the zero-noise ( $\beta \xrightarrow{} \infty$) limit. The zero-noise limit concerns us since any and all randomization in strategic choice vanishes in that limit. Hence we can compare the Nash equilibrium payoff with the APP in the zero-noise limit, which will give us an understanding of how accurate NEM is in predicting the equilibrium behavior of players. In the $\beta \xrightarrow{} \infty$ limit, (\ref{vaccinationgameaveragepayoff}) leads to,
\begin{equation}
\underset{\beta \xrightarrow{} \infty}{lim}\langle \mathcal{U'} \rangle=
\begin{cases}
-\frac{(1-\frac{2}{R_0})r_s-r_v}{2} & ,r_v>(1-\frac{1}{2R_0})r_s. \\
\frac{r_s}{4R_0} & ,\resizebox{.2 \textwidth}{!} {$(1-\frac{5}{2R_0})r_s<r_v<(1-\frac{1}{2R_0})r_s$}. \\
\frac{(1-\frac{2}{R_0})r_s-r_v}{2} & ,r_v<(1-\frac{5}{2R_0})r_s.
\end{cases}
\end{equation}
NEM predicts that APP in zero-noise limit is payoff of strategy $(V,V)$ for $r_v<(1-\frac{5}{2R_0})$ and $(NV,NV)$ for $r_v>(1-\frac{1}{2R_0})$ respectively as can be seen by looking at the transformed payoff matrix $\mathcal{U'}$~(\ref{transformedChickenmatrix}). Additionally when $(1-\frac{5}{2R_0})<r_v<(1-\frac{1}{2R_0})$, $\underset{\beta \xrightarrow{} \infty}{lim}\langle \mathcal{U'} \rangle$ is the average of payoffs from strategic choices $(NV,V)$ and $(V,NV)$. This gives us a picture where for zero-noise, players in thermodynamic limit alternate between $V$ and $NV$ strategies. From real game payoffs~(\ref{vaccinationmatrix}), the APP in the zero-noise limit as via NEM is,
\begin{equation}
\underset{\beta \xrightarrow{} \infty}{lim}\langle \mathcal{U} \rangle=
\begin{cases}
-(1-\frac{1}{R_0})r_s & ,r_v>(1-\frac{1}{2R_0})r_s. \\
\frac{-r_v-r_s+2r_s/R_0}{2} & ,\resizebox{.2 \textwidth}{!} {$(1-\frac{5}{2R_0})r_s<r_v<(1-\frac{1}{2R_0})r_s$}. \\
-r_v & ,r_v<(1-\frac{5}{2R_0})r_s.
\end{cases}
\end{equation}
This coincides with the result, when taking the $\beta \xrightarrow{} \infty$ limit for game magnetization $m_g$. We have plotted the APP in zero-noise limit against vaccination cost $r_v$ in Fig.~(\ref{fig:AvgUvsr1}). These agree with pure strategy Nash equilibrium payoff of vaccination dilemma as in section~\ref{vaccinationdilemma theory} to a large extent. There is however a small deviation where analytical values don't agree with Nash equilibrium payoffs near the domains where Nash equilibrium switches from pure strategy to mixed strategy and vice versa.

\subsection{\label{Chicken-ABM}Results from agent based method}
\subsubsection{\label{Chicken-ABM-Mag}Game magnetization}
\begin{figure}
\centering
\includegraphics[height=6.0cm]{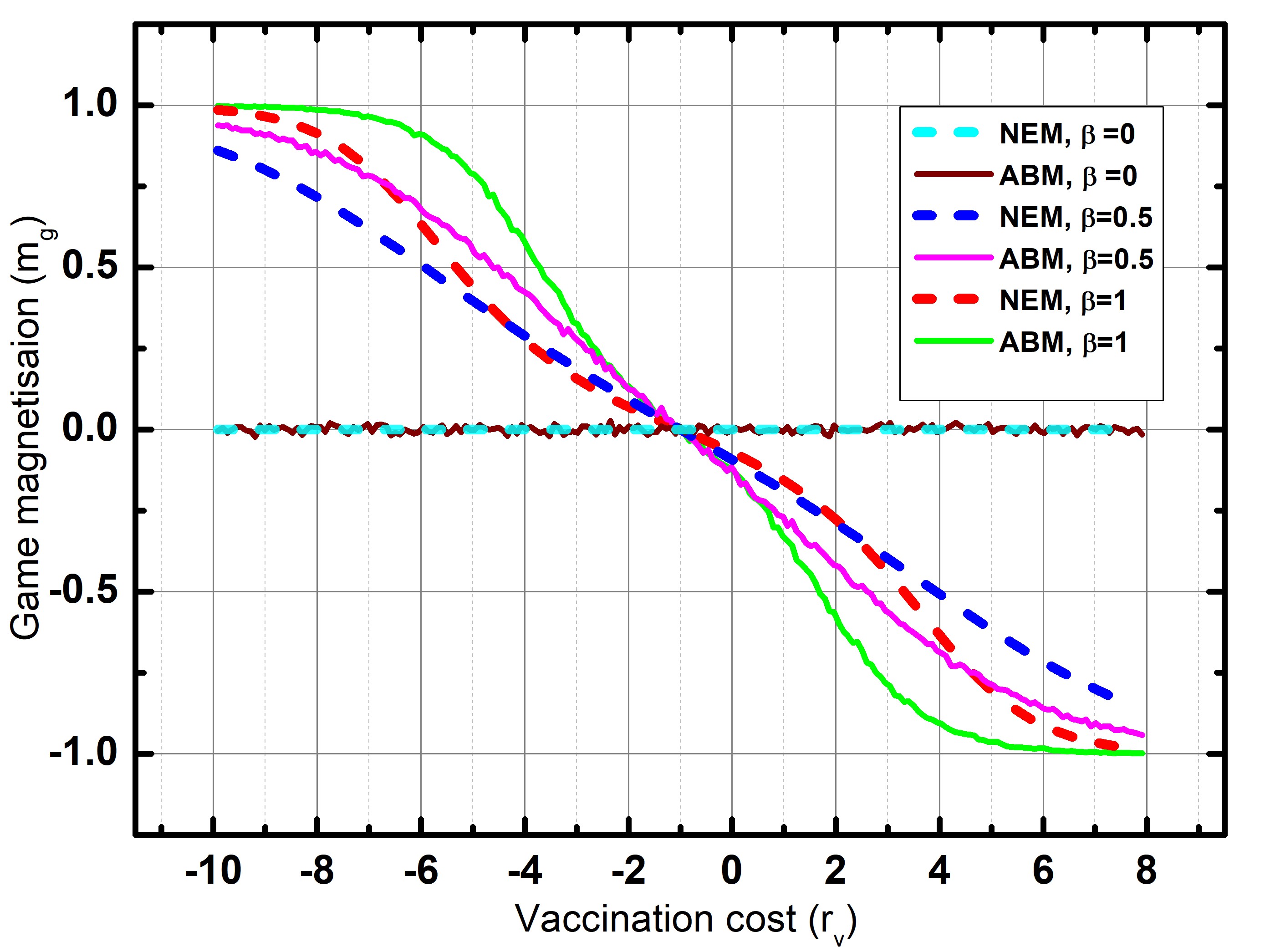}
\caption{How do NEM and ABM compare? $m_g$ vs. vaccination cost ($r_v$) for the vaccination dilemma for cost of disease contraction ($r_s$) $=5$ and $R_0=1.25$.}
\label{fig:Magvsr1}
\end{figure}

For agent based simulation, we have the energy matrix $\mathcal{E}$ as negative of game payoff matrix~(\ref{vaccinationmatrix}) (since players seek the largest payoff and spins seek lowest energy) and follow the algorithm detailed in Sec.~\ref{ABMtheory} to compute game magnetization.
\begin{equation}
\text{Thus, }\mathcal{E}=
\left(\begin{array}{c c}
r_v &r_v\\
(1-\frac{2}{R_0})r_s & (1-\frac{1}{R_0})r_s
\end{array}\right).
\label{eqn:ChickenAgentenergymatrix}
\end{equation}
The payoffs written here are only that of the row player. One can deduce the column player payoffs since the game is symmetric. Fig.~\ref{fig:Magvsr1} shows the game magnetization plotted against vaccination cost $r_v$ computed using this method.
\subsubsection{\label{Chicken-ABM-avgU}Average-payoff/player(APP)}
\begin{figure}
\centering
\includegraphics[height=6.0cm]{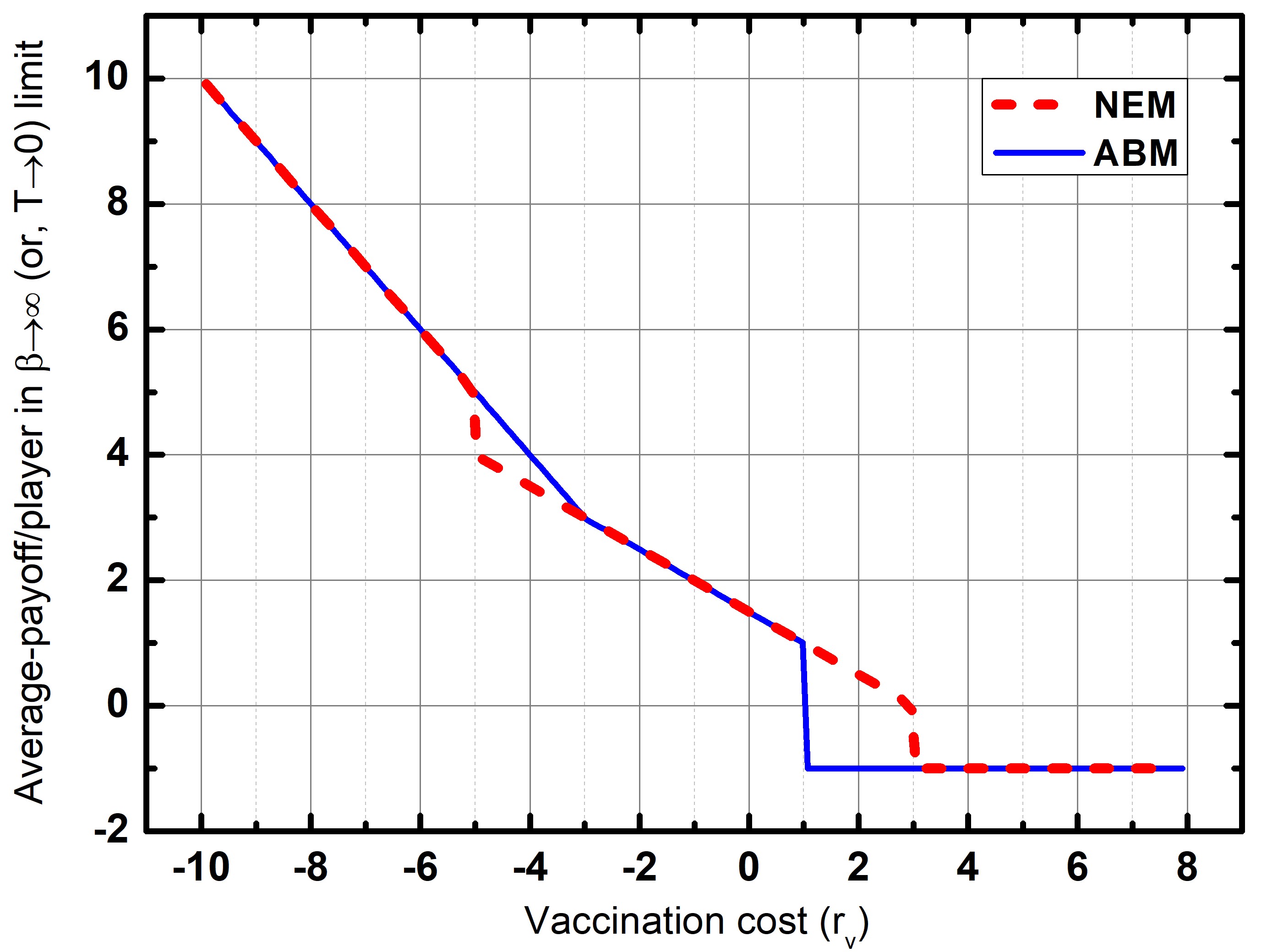}
\caption{ {How do NEM and ABM compare? APP vs vaccination cost ($r_v$) for vaccination dilemma for cost of disease contraction ($r_s$) $=5.0$ and $R_0=1.25$. }}
\label{fig:AvgUvsr1}
\end{figure}

On top of game magnetization, we also calculate the APP in the zero-noise limit. We use the same algorithm as above except that $\beta$ is set at a huge value (say, $\beta=10^6$), and towards the end, we take the average energies of all spin sites in the final state of the system. APP is negative of the average energy since we had started with taking energy matrix $\mathcal{E}$ as negative of payoff matrix $\mathcal{U}$. Fig.~\ref{fig:AvgUvsr1} shows the APP in zero-noise limit plotted against Vaccination cost $r_v$ computed using this method.

\section{Analysis}
We see the game magnetization plotted against $r_v$ in Fig.~\ref{fig:Magvsr1} has much the same trend as the ABM simulation (the Python 3 code for ABM simulation is provided in Appendix B). At the point corresponding to mixed Nash equilibrium where $V$ and $NV$ strategies are equi-probable, we see that game magnetization according to NEM as well as ABM vanishes. When $r_v$ is farther away from this equi-probable mixed strategy point, i.e., payoff of $(1-\frac{3}{2R_0})r_s$, the difference between the two model widens. However for large difference between $r_v$ and $(1-\frac{3}{2R_0})$, both NEM and ABM agree reasonably. For ABM simulation, as $\beta$ increases, $m_g \rightarrow 0$ for payoff values $r_s(1-\frac{2}{R_0})<r_v<r_s(1-\frac{1}{R_0})$, while $m_g \rightarrow +1$ when payoffs obey the condition: $r_v<r_s(1-\frac{2}{R_0})$ and $m_g \rightarrow -1$ when payoffs obey: $r_v>r_s(1-\frac{1}{R_0})$. This is because at larger values of $\beta$, there is less randomization in strategies and more players follow the Nash equilibrium strategy. For NEM, as $\beta$ increases, $m_g \rightarrow 0$ when payoffs obey $r_s(1-\frac{5}{2R_0})<r_v<r_s(1-\frac{1}{2R_0})$, $m_g \rightarrow +1$ for payoffs: $r_v<r_s(1-\frac{5}{2R_0})$ and $m_g \rightarrow -1$ for payoffs: $r_v>r_s(1-\frac{1}{2R_0})$. Hence NEM follows similar behavior to that of ABM simulation, however near payoff domain where Nash equilibrium switches between pure to mixed, there is a slight disagreement between the two.
\begin{table*}[]
\begin{adjustbox}{center}
\begin{tabular}{ccccc|c|c|c|c|c|c|c|c|c|c|}
\cline{6-15}
& & & & & \multicolumn{10}{c|}{$R_0$=1.25} \\ \cline{6-15}
& & & & & \multicolumn{5}{c|}{$m_g$(NEM)} & \multicolumn{5}{c|}{$m_g$(ABM)} \\ \hline
\multicolumn{1}{|c|}{Age group} & \multicolumn{1}{c|}{$r_v^{b}$} & \multicolumn{1}{c|}{$r_v^{h}$} & \multicolumn{1}{c|}{$r_v$} & $r_s$ & $\beta$=0 & $\beta$=0.1 & $\beta$=0.25 & $\beta$=0.5 & $\beta$=1 & $\beta$=0 & $\beta$=0.1 & $\beta$=0.25 & $\beta$=0.5 & $\beta$=1 \\ \hline
\multicolumn{1}{|c|}{20-29} & \multicolumn{1}{c|}{-6.90} & \multicolumn{1}{c|}{1.10} & \multicolumn{1}{c|}{-5.80} & 0.16 & 0.00 & 0.28 & 0.61 & 0.89 & 0.99 & 0.00 & 0.28 & 0.61 & 0.89 & 0.99 \\ \hline
\multicolumn{1}{|c|}{30-39} & \multicolumn{1}{c|}{-24.90} & \multicolumn{1}{c|}{0.80} & \multicolumn{1}{c|}{-24.10} & 0.66 & 0.00 & 0.83 & 0.99 & 1.00 & 1.00 & 0.00 & 0.83 & 0.99 & 1.00 & 1.00 \\ \hline
\multicolumn{1}{|c|}{40-49} & \multicolumn{1}{c|}{-51.50} & \multicolumn{1}{c|}{0.50} & \multicolumn{1}{c|}{-51.00} & 1.99 & 0.00 & 0.99 & 1.00 & 1.00 & 1.00 & 0.00 & 0.99 & 1.00 & 1.00 & 1.00 \\ \hline
\multicolumn{1}{|c|}{50-59} & \multicolumn{1}{c|}{-95.60} & \multicolumn{1}{c|}{0.40} & \multicolumn{1}{c|}{-95.20} & 6.39 & 0.00 & 1.00 & 1.00 & 1.00 & 1.00 & 0.00 & 1.00 & 1.00 & 1.00 & 1.00 \\ \hline
\multicolumn{1}{|c|}{60-69} & \multicolumn{1}{c|}{-127.20} & \multicolumn{1}{c|}{0.20} & \multicolumn{1}{c|}{-127.00} & 13.92 & 0.00 & 1.00 & 1.00 & 1.00 & 1.00 & 0.00 & 1.00 & 1.00 & 1.00 & 1.00 \\ \hline
\end{tabular}
\end{adjustbox}
\caption{Game magnetization $m_g$ as obtained by Nash equilibrium mapping (NEM) and agent based method (ABM) for $R_0=1.25$ at $\beta=0.25,0.5,1.0$. $r_v$ values are obtained for different age groups by adding the potential benefits $r_v^b$(prevented ICU admissions per 100,000 population in UK) and harm $r_v^h$ (reported cases of blood clots per 100,000 population in UK) caused by Astra-Zeneca COVID-19 vaccine (according to data provided by UK Winton Centre for Risk and Evidence Communication\cite{Miller}). $r_s$ values are obtained for different age groups by and death counts due to COVID-19 (per 100,000 population from England and Wales, according to data provided by the Office of National statistics, UK \cite{ONSdeaths}).}
\label{table:table1}
\end{table*}

APP in zero-noise limit, plotted against $r_v$ in Fig.~\ref{fig:AvgUvsr1} matches with agent based method for a fairly large domain near the equi-probable mixed strategy Nash equilibrium. However, for payoffs in the range domains $r_s(1-\frac{5}{2R_0})<r_v<r_s(1-\frac{2}{R_0})$ and $r_s(1-\frac{1}{R_0})<r_v<r_s(1-\frac{1}{2R_0})$, i.e, near the domains where Nash equilibrium of the game switches between pure and mixed, the APP from NEM does not agree with that of ABM simulation.  On the other hand, ABM gives payoffs corresponding to $(V,V)$ for $r_v<r_s(1-\frac{2}{R_0})$, $(NV,NV)$ for $r_v>r_s(1-\frac{1}{R_0})$ and the average of $(V,NV)$ and $(NV,V)$ for $r_s(1-\frac{1}{R_0})<r_v<r_s(1-\frac{2}{R_0})$- all corresponding to the Nash equilibrium in their respective domains. Hence, NEM agrees with ABM except in the small payoff domain $r_s(1-\frac{5}{2R_0})<r_v<r_s(1-\frac{2}{R_0})$ and $r_s(1-\frac{1}{R_0})<r_v<r_s(1-\frac{1}{2R_0})$.

\section{Comparison with real-life data from Astra-Zeneca COVID-19 vaccine rollout in UK }
Table~\ref{table:table1} lists some magnetization values from real-life data for $r_v$ and $r_s$. For $r_v$, we consider the potential benefits of Astra-Zeneca COVID-19 vaccine $r_v^b$ (avoided COVID-19 ICU admissions per 100,000 population) and add the measure of harm caused by the vaccine $r_v^h$(reported blood clot incidents per 100,000 population). UK Winton Centre gathered this data for Risk and Evidence Communication collected since the vaccination roll-out began on Dec 8, 2020, up until Apr 28, 2021, based on COVID-19 Infection Survey, ONS Apr 30 2021\cite{Miller, survey}. Since $r_v$ had been defined as the cost incurred by an individual, $r_v^b$ will be a negative value, and $r_v^h$ will be positive. $r_s$ is the cost of disease contraction, and these are taken as the deaths occurred in England and Wales from data collected since Dec 8, 2020, up until Apr 28, 2021 \cite{ONSdeaths}. We specifically chose the death counts as a measure of $r_s$ since the general populace is motivated to take up vaccines mostly by looking at disease mortality rates. In Table~\ref{table:table1}, we have $m_g$ values (which gives the difference between the fraction of vaccinated and unvaccinated individuals within a population) obtained for different age groups at $R_0=1.25$ using NEM and ABM simulation. $R_0$ value was chosen as such based on the average weekly estimate of $R_0$ of COVID-19 recorded in England during Dec 8, 2020, and Apr 30, 2021 \cite{rvalue}. We see that for all age groups, $m_g$ approaches $+1$ as $\beta$ increases. This happens because $r_v<r_s(1-\frac{2}{R_0})$ for all values of $r_v$ and $r_s$ in Table.~\ref{table:table1}. It means the Nash equilibrium of the game is $(V, V)$ for all parameters in Table.~\ref{table:table1}. Therefore, as $\beta$ increases, randomness in choice of strategy decreases, and more players end up choosing vaccination strategy. Small $\beta$ implies people have more choice and agency whether to take the vaccine or not. Larger $\beta$ means larger the fraction population adopting Nash equilibrium strategy, i.e., Vaccination. Hence, $\beta$ acts as a measure of government intervention regarding vaccination mandates and media propaganda. $\beta=0$ means people are left on their own will, and no amount of awareness or enforcement is happening on behalf of government or media. $\beta \ge 0.5$ would mean government giving incentives to take up vaccination  while $\beta \ge 1.0$ would mean government makes vaccines mandatory and $\beta \le 0.5$ would mean only media is making noises and there is no government intervention as such. For the basic reproductive ratio value of $R_0 = 1.25$ and $\beta = 0.1,0.25,0.5$ and $1.0$, we see nearly everyone above the age of $40$ showing nearly full commitment towards taking vaccines since the magnetization value is equal to (or extremely close) to $+1$, meaning 100$\%$ of the population is vaccinated. Individuals between ages $30-39$ show a small reluctance compared to their older counterparts at smaller $\beta$ values. For example, at $\beta = 0.1$, we have $m_g = 0.83$, while for  $\beta = 0.25$, we have $m_g = 0.99$. Thus, for the age group $30-39$, around $91.5\%$ of the population takes up vaccination at $\beta=0.1$ and for $\beta=0.25$, this age group reaches full vaccination threshold. This means for this age group only media campaigns would be enough to get the population vaccinated. On the other hand, the population belonging to the $20-29$ age group shows a reluctance to get vaccinated. That means for the age group ($20-29$) vaccination mandates have to be enforced by the government. However, for the same parameters, every other age group is nearly fully vaccinated. It happens because the relative vaccine-benefit to disease-cost ratio is smaller for young people ($20-39$) than older individuals($40-69$).  It is further corroborated by Ref.~\cite{Jhun}, which also studied epidemic spread in a meta-population divided into different age groups. The study, using numerical simulations, found that in case of the limited supply of vaccines and for large reproductive ratio $R_0$ (greater than $1.12$), the optimal strategy for decreasing the death rate is to vaccinate the population from oldest to youngest in the descending order. Hence, our projection predicts that efficient vaccination coverage among younger age groups would require greater effort from the government and media in terms of awareness than their older counterparts.

\section{Conclusion}
Our analysis of the vaccination dilemma in thermodynamic limit shows us the three domains which differ in terms of Nash equilibrium- first where $(V, V)$ is the equilibrium, next with $(NV, NV)$ being Nash equilibrium, and finally with $(V, NV)$,$(NV, V)$ being Nash equilibria along with a mixed Nash equilibrium where players choose $V$ strategy with probability $R_0(1-\frac{1}{R_0}-\frac{r_v}{r_s})$. Game magnetization and APP in zero noise limit obtained via NE mapping seem to agree reasonably with ABM simulations and the above-mentioned Nash equilibrium of the Vaccination dilemma. However, near the points where Nash equilibrium of the game switches from pure to mixed, there is a slight disagreement between the methods, i.e., in the payoff range $r_s(1-\frac{5}{2R_0})<r_v<r_s(1-\frac{2}{R_0})$ and $r_s(1-\frac{1}{R_0})<r_v<r_s(1-\frac{1}{2R_0})$. Additionally, analyzing the Astra-Zeneca COVID-19 vaccine risk and disease deaths, both NE mapping and agent-based methods predict near $100\%$ AZ vaccine coverage above age $40$. At the same time, younger people show a slight  reluctance due to the relative risk to benefit quotient being non-negligible. $\beta$ acts as a numerical measure of government intervention via advertisement campaigns, etc,  to get vaccinated. Older age groups ($40-69$) getting close to $100\%$ Vaccination at small $\beta$ values indicates government intervention isn't  necessary to get full coverage. In contrast, for the younger population ($20-39$), some encouragement from the government via media campaigns may be necessary. We also should mention a few limitations of our study: firstly, this is only for Astra-Zeneca vaccine, and secondly, the data is from UK only on blood clots. So it should not be taken to mean all vaccines would give similar results. Further, for Astra-Zeneca itself, results in different countries may be different. It will be nice to extend this analysis to other vaccines like Pfizer, Moderna, Johnson and Johnson, Sputnik, Sinovac and Covaxin vaccine. Further, going beyond vaccination strategies, we would also like to do a deep dive into lock-down measures\cite{cheong} to prevent spread of COVID-19 in a game theoretic setting, via the Nash equilibrium mapping method\cite{NEvsDE}. 
\section*{Competing Interests} The authors declare no competing interests.
\acknowledgments Colin Benjamin would like to thank Science and Engineering Research Board (SERB) for funding under the Core Research grant "Josephson junctions with strained Dirac materials and their application in quantum information processing," Grant No. CRG/2019/006258.
\section{Appendix}
\begin{appendix}
This Appendix has two sections, first deals with the problems associated with another analytical method, namely Darwinian evolution. In the second section we provide a Python 3 code for Agent based method simulation.
\section{\label{DEtheory}Darwinian evolution (DE) model}
Much like NE mapping, there are different analytical methods introduced in a previous paper (see Ref.~\cite{Adami}) that uses analogy with $1D$ Ising model to derive formulation for game magnetization. Hamiltonian dynamics (HD) model one such method. For a game matrix that obeys the condition $a_{1}+a_{4}=a_{2}+a_{3}$, the game magnetization according to HD model is,
\begin{align}
m_g^{HD}=\tanh\beta\left( \frac{a-d}{2}\right).
\label{magHD}
\end{align}
However, as we can see vaccination dilemma does not obey the $a_{1}+a_{4}=a_{2}+a_{3}$ payoff condition and hence HD model is not applicable. HD model game magnetization does not reduce to a concise analytical form unless this condition is true. Ref.~\cite{Adami} also introduced another model called Darwinian evolution (DE) to calculate the game magnetization, which produced modestly improved results compared to the HD model. DE model can be applied for a general symmetric payoff matrix, and hence it is not limited to games that satisfy the condition on payoffs $a_{1}+a_{4}=a_{2}+a_{3}$ for payoffs. The analytical foundation of the DE model is the same as that of the HD model, except the Hamiltonian of the system of spin sites/players is chosen. It focuses only on the energy of a single spin or player. The spin site under focus is with the $i=1$ index. Hamiltonian for DE model is defined such that it represents the energy of a single spin interacting with its nearest neighbor in a 1D lattice. It is given by,
\begin{equation}
\mathcal{H}_1=\sum_{m,n=0}^1 \mathcal{E}_{mn}P_m^{(1)}\otimes P_n^{(2)}.
\end{equation}
Here the energy matrix is related to the game payoffs as,
\begin{equation}
\mathcal{E}=\begin{pmatrix}
\mathcal{E}_{00} & \mathcal{E}_{01}\\
\mathcal{E}_{10} & \mathcal{E}_{11}
\end{pmatrix}
=\begin{pmatrix}
-a_{1} & -a_{2}\\
-a_{3} & -a_{4}
\end{pmatrix}.
\label{payoffandenergymatrix}
\end{equation}
DE model minimizes the energy of the spin site under focus (i.e., it maximizes the payoff of a single player). Therefore for a $1D$ Ising system consisting of two spin sites, it would mean minimizing the energy (or maximizing the payoff) of site $1$ while considering its interaction with site $2$. $\ket{\alpha}=\ket{m_1m_2}$ represents the state of a $1D$ lattice with $2$-spin sites. The partition function for such a system according to the DE model is as follows:
\begin{align}
\mathcal{Z}=&\sum_{\ket{\alpha}}\bra{\alpha}e^{-\beta \mathcal{H}_1}\ket{\alpha}=\sum_{m_1m_2}\bra{m_1m_2}e^{-\beta \mathcal{H}_1}\ket{m_1m_2}\\
=&e^{\beta a_{1}} + e^{\beta a_{2}} + e^{\beta a_{3}} + e^{\beta a_{4}}.
\label{eqn:DEpartitionfunction}
\end{align}
DE model cares only for magnetization produced by spin site $1$, hence we have introduce selective order parameter $\hat{M_z}^{(1)}=P_0^{(1)}-P_1^{(1)}$ (See Refs.~\cite{Adami,NEvsDE} for details). The game magnetization is then given as,
\begin{align}
m_g^{DE}=\langle \hat{M_z}^{(1)}\rangle_\beta=&\frac{1}{\mathcal{Z}}\sum_{\ket{\alpha}}\bra{\alpha} \hat{M_z}^{(1)}e^{-\beta \mathcal{H}_1}\ket{\alpha}\\
=&\frac{e^{\beta a_{1}} + e^{\beta a_{2}} -e^{\beta a_{3}} - e^{\beta a_{4}}}{e^{\beta a_{1}} + e^{\beta a_{2}} + e^{\beta a_{3}} + e^{\beta a_{4}}}.
\label{eqn:MagDEgeneral}
\end{align}
From Eq.~(\ref{vaccinationmatrix}), we get the game magnetization for vaccination game as,
\begin{equation}
m_g^{DE}=\frac{2e^{-\beta r_v} - e^{-\beta r_s}(e^{\beta r_s/R_0}+e^{2\beta r_s/R_0})}{2e^{-\beta r_v} + e^{-\beta r_s}(e^{\beta r_s/R_0}+e^{2\beta r_s/R_0})}.
\end{equation}
\begin{figure}
\centering
\includegraphics[height=6.0cm]{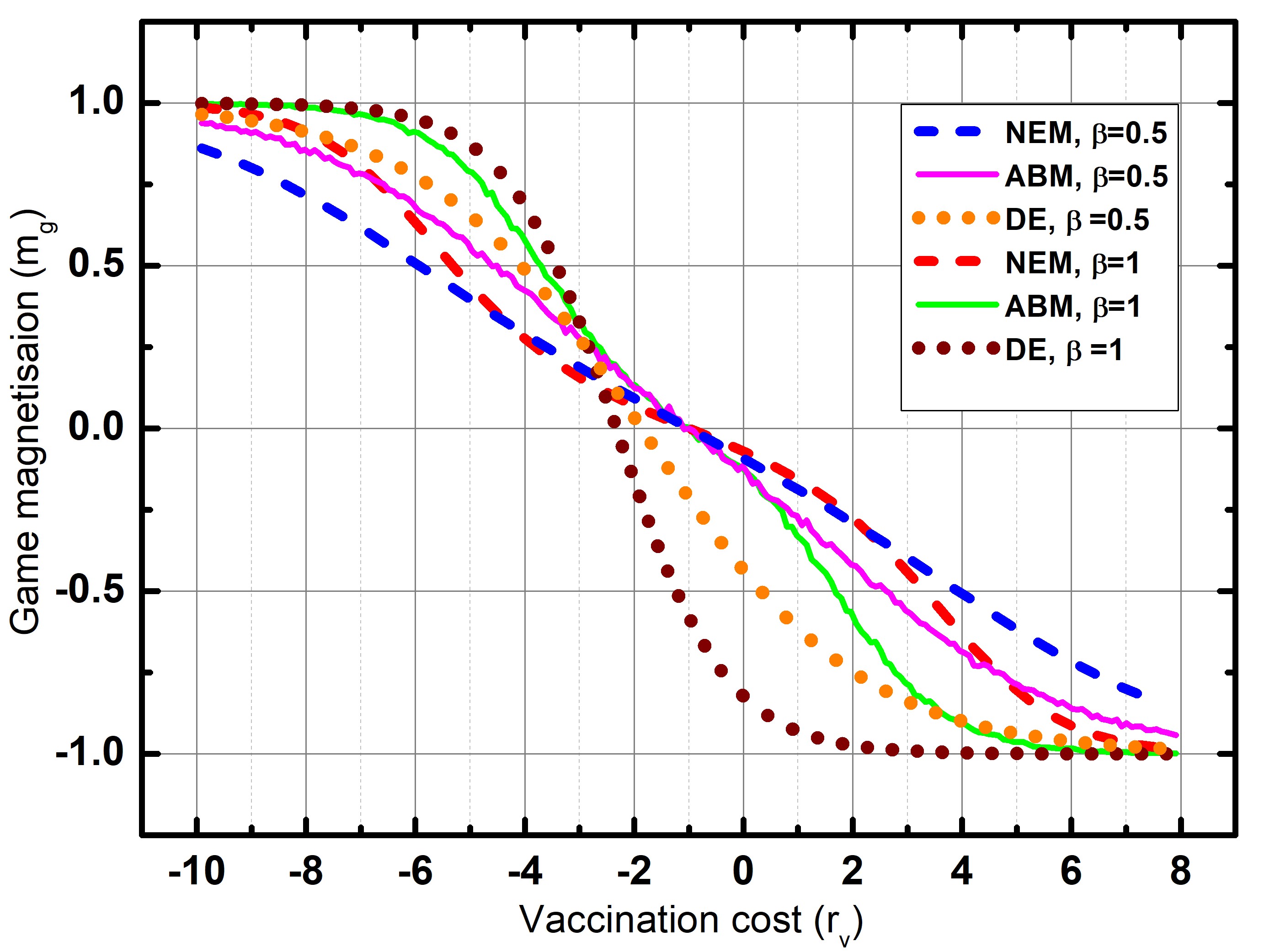}
\caption{How do NEM, ABM and DE compare? $m_g$ vs vaccination cost ($r_v$) for the vaccination dilemma, with cost of disease contraction ($r_s$) $=5.0$ and $R_0=1.25$. }
\label{fig:MagvsrwithDE}
\end{figure}
Game magnetization $m_g$ is plotted against $r_v$ in Fig~\ref{fig:MagvsrwithDE} as generated by NEM, DE model and ABM. We see that the DE model differs a great deal from NEM  and ABM. Additionally, $m_g$ derived via the DE model vanishes at a payoff value that does not correspond to the payoff where $m_g$ vanishes for NEM (and ABM) and is hence non-representative the mixed Nash equilibrium of the vaccination game. Although in DE model $m_g$ seems to agree with agent-based results for some values of payoff $r_v<r_s(1-\frac{2}{R_0})$, their disparity becomes quite significant for $r_v>(1-\frac{2}{R_0})$. It happens because the DE model is concerned with the payoff of a single player in the entire chain and fails to account for all the players' collective Nash equilibrium. The strategy corresponding to the largest payoff in the payoff matrix also happens to be the Nash equilibrium strategy when $r_v<r_s(1-\frac{2}{R_0})$. However, for $r_v>(1-\frac{2}{R_0})$, Nash equilibrium strategy and largest payoff strategy is no longer the same and hence $m_g$ values for both models diverge. We will further see the confirmation of this when we look at the APP.

For  APP, we start with the partition function~(\ref{eqn:DEpartitionfunction}) for vaccination dilemma which can be obtained as,
\begin{equation}
\mathcal{Z}=2e^{-\beta r_v} + e^{-\beta r_s}(e^{\beta r_s/R_0}+e^{2\beta r_s/R_0}).
\end{equation}
We find average thermodynamic energy $\langle \mathcal{E} \rangle$ using~(\ref{eqn:averagethermodynamicenergy}). APP $\langle \mathcal{U} \rangle$ is given by negative of this value. Unlike NEM, we do not halve it since energy considerations were limited to a single spin site from the beginning.
\begin{align}
&\langle \mathcal{U} \rangle=-\langle \mathcal{E} \rangle=\frac{\partial \ln \mathcal{Z}}{\partial \beta}=\\
&-\frac{\resizebox{.45 \textwidth}{!} {$2r_ve^{-\beta r_v}+r_s(1-\frac{1}{R_0})e^{-\beta r_s(1-\frac{1}{R_0})}+r_s(1-\frac{2}{R_0})e^{-\beta r_s(1-\frac{2}{R_0})}$}}{2e^{-\beta r_v}+e^{-\beta r_s(1-\frac{1}{R_0})}+e^{-\beta r_s(1-\frac{2}{R_0})}}.
\end{align}
APP in the infinite-noise limit can be obtained as,
\begin{equation}
\underset{\beta \xrightarrow{} 0}{lim} \langle \mathcal{U} \rangle = \frac{-r_v-r_s+3r_s/2R_0}{2}.
\end{equation}

\begin{figure}
\centering
\includegraphics[height=6.0cm]{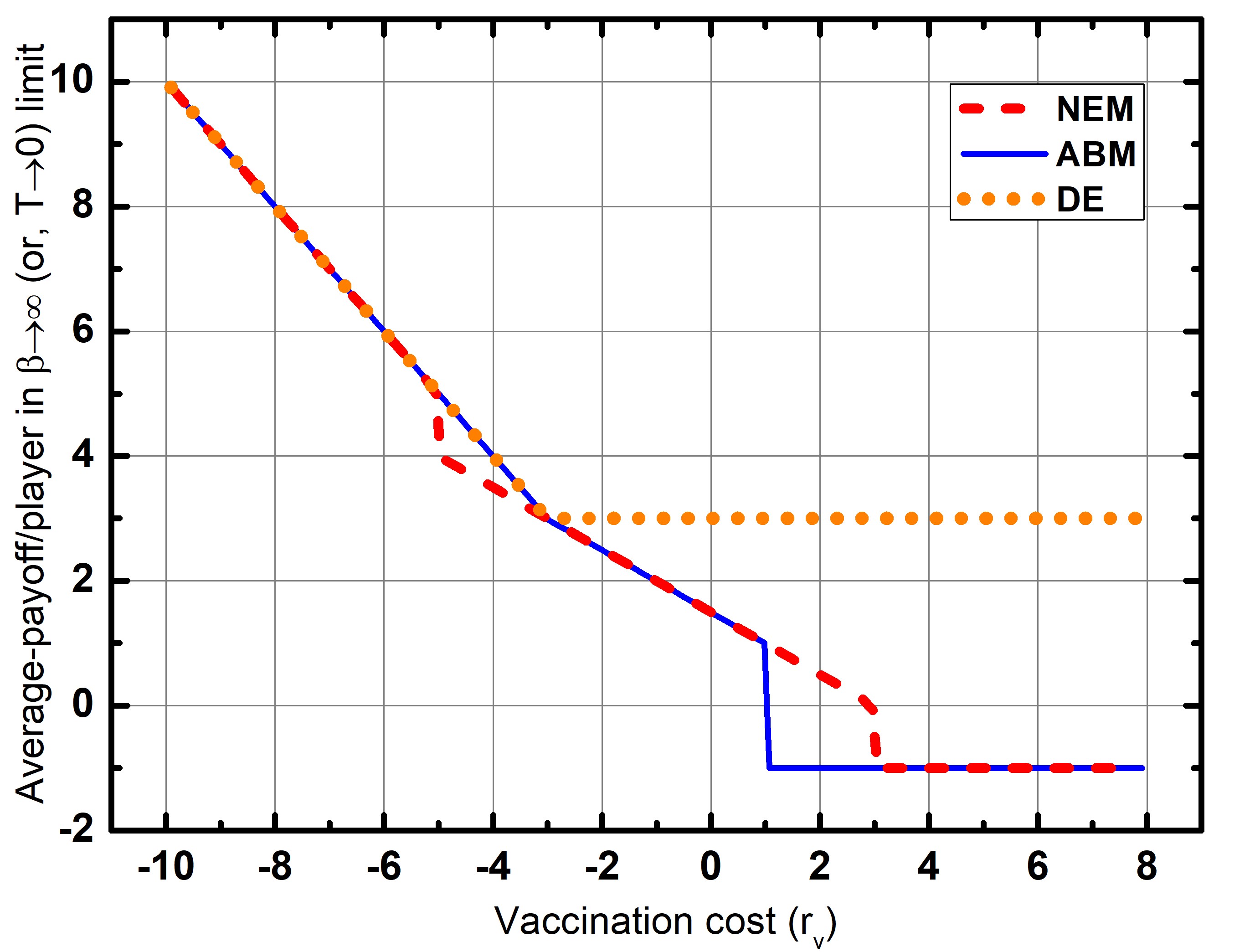}
\caption{ How do NEM, ABM and DE compare? APP vs vaccination cost ($r_v$) for the vaccination dilemma with  cost of disease contraction ($r_s$) $=5.0$ and $R_0=1.25$. }
\label{fig:AVGUvsrWITHDE}
\end{figure}

This result is same as that obtained via NEM, see~(\ref{zeroUlimit}). This is the result for all strategy pairs being equally probable in the infinite-noise limit. Calculating APP in zero-noise limit results in,
\begin{equation}
\underset{\beta \xrightarrow{} \infty}{lim}\langle \mathcal{U} \rangle=
\begin{cases}
-r_v & ,r_v<(1-\frac{2}{R_0})r_s. \\
-(1-\frac{2}{R_0})r_s & ,r_v>(1-\frac{2}{R_0})r_s.
\end{cases}
\end{equation}

APP in the zero-noise limit has been plotted against $r_v$ in Fig.~\ref{fig:AVGUvsrWITHDE}. We can see that the DE model differs substantially from both ABM and NEM. Similar to our observation from game magnetization, DE model and ABM agree for $r_v<r_s(1-\frac{2}{R_0})$ but they give different outcomes for $r_v>(1-\frac{2}{R_0})$. Here we can see a clearer picture as the APP in the range $r_v>r_s(1-\frac{2}{R_0})$ is equal to $-r_v$, which happens to be the largest payoff in that range. However, for $r_v>r_s(1-\frac{2}{R_0})$, the largest payoff in the payoff matrix is $(1-\frac{2}{R_0})$ which isn't the payoff associated with the Nash equilibrium and hence DE model and ABM results diverge.
\onecolumngrid
\section{Python code to calculate game magnetization of vaccination dilemma via numerical agent based method}
The python3 code we used for finding the game magnetization vs vaccination cost ($r_v$) graph for the Vaccination game (see FIG. 1) using Agent-based simulation is given below.
\lstset{language=Python}
\lstset{frame=lines}
\lstset{label={lst:code_direct}}
\lstset{basicstyle=\footnotesize}
\begin{lstlisting}
import numpy as np 
import matplotlib.pyplot as plt 
a=np.random.rand(1000) 
#1-D string of 10,000 players 
PU=np.linspace (-10.0,8.0,200) 
#Domain of punishment where we plot magnetisation 
for i in range(0,len(a)): 
    if(a[i]<0.5):
        a[i]=int('0')
    else:
        a[i]=int('1') 
m=[] 
R0=1.25 #temptation payoff 
rs=5.0 #reward payoff 
T=1.0 #Temperature 
for rv in PU:
    E=[[rv,rv],[ (1-2/R0)*rs, (1-1/R0)*rs]] #Energy matrix
    for k in range(0,1000000): 
    #10 million iterations; average 1000 iterations per player
        i=np.random.randint(len(a))
     #Randomly choosing a player 
        p=np.random.rand() 
     #Random value of p between 0 and 1
        if((p)<=(1/(1+np.exp(-(E[int(a[i])][int(a[(i+1)\%len(a)])]
        -E[int(a[i]+1)\%2][int(a[(i+1)\%len(a)])])/T)))): 
            a[i] = (int(a[i]+1)\%2) 
        #Flipping the strategy when p< 1/(1+e\beta\Delta E) 
    m=np.append(m,(len(a)-2*sum(a))*1.0/len(a)) 
#Magnetisation values 
plt.plot(PU,m,label='Agent based model')
\end{lstlisting}

\end{appendix}

\end{document}